\documentclass[conference]{IEEEtran}
\IEEEoverridecommandlockouts

\usepackage{cite}
\usepackage{amsmath,amssymb,amsfonts}
\usepackage{algorithmic}
\usepackage{graphicx}
\usepackage{textcomp}
\usepackage{xcolor}

\def\BibTeX{{\rm B\kern-.05em{\sc i\kern-.025em b}\kern-.08em
    T\kern-.1667em\lower.7ex\hbox{E}\kern-.125emX}}
\begin{document}

\title{SkyTrust: Blockchain-Enhanced UAV Security for NTNs with Dynamic Trust and Energy-Aware Consensus\\

\thanks{Identify applicable funding agency here. If none, delete this.}
}

\author{\IEEEauthorblockN{ Afan Ali}
\IEEEauthorblockA{\textit{School of Engineering and Natural Sciences} \\
\textit{Istanbul Medipol University} \\
Istanbul, Turkey \\
afanali85@gmail.com}
\and
\IEEEauthorblockN{Irfanullah Khan Shahbaz }
\IEEEauthorblockA{\textit{Interdisciplinary Research Center for Smart Mobility and Logistics} \\
\textit{King Fahd University of Petroleum and Minerals}\\
Dhahran, Saudi Arabia \\
irfanullah.khan@kfupm.edu.sa}
}

\maketitle

\begin{abstract}
Non-Terrestrial Networks (NTNs) based on Unmanned Aerial Vehicles (UAVs) as base stations are extremely susceptible to security attacks due to their distributed and dynamic nature, which makes them vulnerable to rogue nodes. In this paper, a new Dynamic Trust Score Adjustment Mechanism with Energy-Aware Consensus (DTSAM-EAC) is proposed to enhance security in UAV-based NTNs. The proposed framework integrates a permissioned Hyperledger Fabric blockchain with Federated Learning (FL) to support privacy-preserving trust evaluation. Trust ratings are updated continuously through weighted aggregation of past trust, present behavior, and energy contribution, thus making the system adaptive to changing network conditions. An energy-aware consensus mechanism prioritizes UAVs with greater available energy for block validation, ensuring efficient use of resources under resource-constrained environments. FL aggregation with trust-weighting further increases the resilience of the global trust model. Simulation results verify the designed framework achieves 94\% trust score prediction accuracy and 96\% rogue UAV detection rate while outperforming centralized and static baselines of trust-based solutions on privacy, energy efficiency, and reliability. It complies with 6G requirements in terms of distributed intelligence and sustainability and is an energy-efficient and scalable solution to secure NTNs.
\end{abstract}

\begin{IEEEkeywords}
Non-Terrestrial Networks, UAV Security, Blockchain, Federated Learning, Dynamic Trust Adjustment, Energy-Aware Consensus.
\end{IEEEkeywords}

\section{Introduction}

\begin{figure}[htbp]
\centerline{\includegraphics[width=8cm]{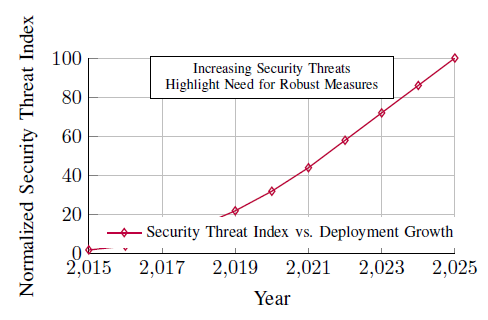}}
\caption{Trends in NTN Ecosystem from 2015 to 2025, showing a combined normalized curve of security incidents and UAV deployment, emphasizing the importance of security as NTN adoption grows.}
\label{graph-uav}
\end{figure}

Deployment of Unmanned Aerial Vehicles (UAVs) as Internet of Things (IoT) or flying Base Stations (BS) devices represent the main paradigm for future communication systems.This has raised the limits for expanding connectivity via conventional terrestrial networks. In all these instances, UAVs are centralized ground-based base stations (BS) or processing units (PU) and are networked in an attempt to offer data exchange as well as coordination of operations. But the dynamic and distributed nature of such Non Terrestrial Networks (NTN) brings critical security concerns where the threat of rogue UAV nodes entering the network and tampering with the integrity of the whole system is always possible. In \cite{b1}, authors recognize UAV-based NTNs vulnerability to unauthorized entry and exposure of data, and therefore insist on strict security in UAV-to-BS communication. Fig.\ref{graph-uav} gives an overview of trends in security incidents and UAV deployment, emphasizing the importance of security as focus on NTN grows. Similarly, authors in \cite{b2} also define the growing potential of threat being caused by malicious nodes in NTN networks, suggesting decentralized protection methods to enable operating robustness.

As opposed to such attacks, blockchain technology suggests a promising ledger structure that records and authenticates all the transactions within a network, so it becomes transparent and penetration-proof. A blockchain framework can be employed as a security system for protecting UAV nodes in NTNs to record and authenticate all the participating nodes, and hence preventing malicious UAVs from accessing the system. The work in \cite{b3} demonstrates how blockchain can be utilized for enabling data integrity and device authentication in UAV-IoT networks despite some considerable implementation challenges. Likewise, in \cite{b4}, authors demonstrate how blockchain can be harnessed for securing UAV-IoT networks, reflecting on its potential for distributed trust management. However, direct application of blockchain in resource-constrained NTN framework is extremely challenging and prone to high computational overheads and energy-expensive operations. In \cite{b5}, authors observe that traditional blockchain consensus mechanisms, i.e., Proof-of-Work, are inapplicable to IoT due to their energy demands, while authors in \cite{b6} discuss overheads of implementing blockchain on resource-limited IoT devices.

To tackle the security challenges in UAV-based Non-Terrestrial Networks (NTNs), we propose a new Dynamic Trust Score Adjustment Mechanism with Energy-Aware Consensus (DTSAM-EAC). Our approach employs a permissioned Hyperledger Fabric blockchain, with Federated Learning (FL) for ensuring privacy-preserving trust evaluation, and dynamic trust score adjustment based on historical trust, recent behavior, and energy contribution. In contrast to \cite{b7}, which employs static trust models for FL, our dynamic adjustment accommodates evolving UAV behavior to better enable rogue node detection. In contrast to \cite{b9}, which is energy-aware but lacking in trust dynamics, our energy-aware consensus protocol selects UAVs with greater energy reserves for block validation, rendering it more efficient. Trust-weighted FL aggregation improves reliability according to 6G requirements of distributed intelligence and greenness, with scalability and adaptability for NTN security. Thus, in this paper, our proposed method has several novel contributions:
\begin{itemize}
\item Unlike centralized trust evaluation methods, our approach combines FL with a Hyperledger Fabric blockchain to train and aggregate trust models in a decentralized and privacy-preserving manner, taking into account the dynamics of NTNs \cite{b7}.
\item  By employing permissioned blockchain with lightweight consensus, we attain energy saving, making the system feasible for energy-constrained UAVs, an improvement over energy-intensive public blockchains \cite{b9}.
\item The system supports 6G requirements of privacy-preserving AI and distributed intelligence, ensuring feasibility for prospective standards.
\item We conduct a novel performance analysis of FL in NTNs, including accuracy, communication overhead, and convergence time, with novel insights into its feasibility under high-mobility conditions \cite{b8}.

\end{itemize}

\section{System Model and Problem Formulation}

\begin{figure}[htbp]
\centerline{\includegraphics[width=8cm]{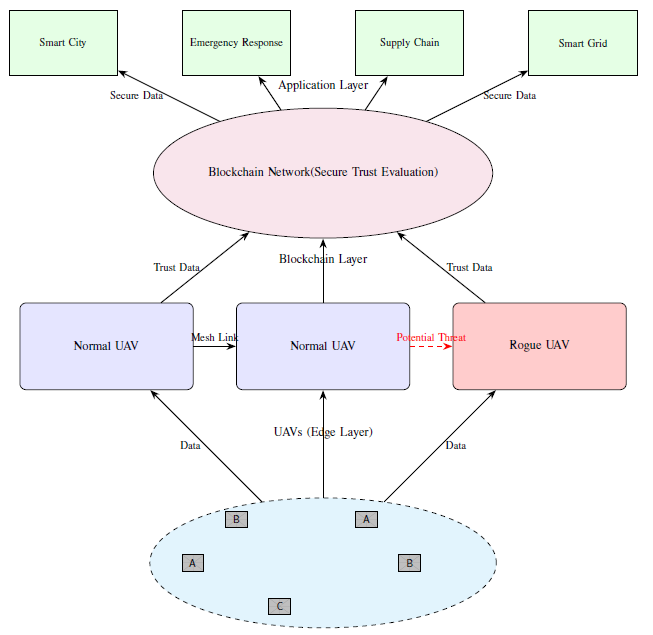}}
\caption{Blockchain based UAV security system for NTNs, showing the integration of ground users, UAVslockchain for secure trust evaluation, and application scenario.}
\label{broad-BD}
\end{figure}

We consider an NTN system incorporating UAVs employed as IoTs or processing units (PUs) to offer connectivity to numerous ground users in dynamic and distributed environments. Fig.\ref{broad-BD} demonstrates such scenario, representing the intercombination of various elements over four conceptual layers. At the ground level, the Device Layer is made up of ground users, represented by a node of IoT devices and smartphones, that generate data and send it to the UAVs in the Edge Layer above them. The Edge Layer is made up of normal UAVs (blue) and a rogue UAV (red) linked by mesh links, which symbolize potential dangers from the rogue UAV. On top of this, the Blockchain Layer is the fundamental security foundation and accepts trust data from the UAVs to securely and tamper-free handle the trust. Finally, the Application Layer shows a few NTN applications and illustrates how the blockchain ensures secure data delivery for these applications. The network is characterized by the following:

\begin{itemize}
    \item \textbf{High Mobility}: UAVs possess high mobility with varying speeds depending on mission requirements (e.g., surveillance, disaster response), which affects link stability and connectivity.
    \item \textbf{Varied Densities}: UAV densities range from rural low deployments to urban high deployments or big events, influencing interference and coverage.
    \item \textbf{Different Topologies}: Topologies of networks are different, e.g., mesh networks with peer-to-peer communications, star topologies with one central UAV, or hierarchical topologies, with diverse security problems.
\end{itemize}

With dynamics as above, the NTN is vulnerable to hostile UAV nodes compromising data integrity or hijacking operations and needs strong security controls.

\subsection{Blockchain and Federated Learning Integration}
In order to address these security needs, we recommend a trust-based access control scheme founded upon a permissioned Hyperledger Fabric blockchain supplemented by Federated Learning (FL) for trust analysis. Hyperledger Fabric delivers trusted, open, and tamper-evident transactions, including a distributed trust activity ledger. Aggregation rules of FL and access trust policy are enforced via smart contracts.

FL enables every UAV to learn a local trust model from their exchanges and observations (e.g., response times, communication patterns), maintaining privacy unbroken without transferring raw data. The FL process involves: (1) each UAV \( u_i \) trains a local model \( f_{\theta_i} \) on its dataset \( D_i \), (2) model parameters \( \theta_i \) are shared via the blockchain, (3) a smart contract aggregates them into global parameters \( \theta \), using:
\[
\theta = \sum_{i=1}^n \frac{|D_i|}{\sum_{j=1}^n |D_j|} \theta_i,
\]
and (4) the global model \( f_\theta \) is distributed back to UAVs. We use TensorFlow Federated for FL implementation, aligning with 6G's emphasis on distributed intelligence and privacy-preserving AI \cite{b7}.

\subsection{Problem Formulation}
Let \(\mathcal{U} = \{u_1, u_2, \ldots, u_n\}\) be the set of UAVs, each maintaining a local dataset \( D_i \) of interaction records (e.g., packet delivery ratio, response times, cooperation metrics). The trust model \( f_\theta \), parameterized by \( \theta \), maps features to a trust score \( T(u_j) \in [0, 1] \), where \( T(u_j) = 1 \) indicates fully trustworthy and \( T(u_j) = 0 \) indicates rogue.

In the FL process, each UAV \( u_i \) trains \( f_{\theta_i} \) on \( D_i \), minimizing a local loss (e.g., cross-entropy). Parameters \( \theta_i \) are shared via the blockchain, aggregated into \( \theta \), and the process iterates for \( T \) rounds until convergence (change in accuracy \( < \epsilon \), e.g., 0.01).

We define performance metrics: (1) \textit{Accuracy in Trust Score Prediction}, as \( \text{Accuracy} = \frac{\text{Correct Predictions}}{\text{Total Predictions}} \), (2) \textit{Communication Overhead}, as \( n \cdot \text{Size}(\theta_i) + \text{Blockchain Overhead} \) per iteration, and (3) \textit{Convergence Time}, as the number of iterations \( T \). These are compared with a centralized approach, where all \( D_i \) are sent to a server, with overhead \( \sum_{i=1}^n |D_i| \), to highlight FL's privacy and efficiency benefits \cite{b8}.

\section{Proposed Solution}

\begin{figure*}[htbp]
\centerline{\includegraphics[width=12cm]{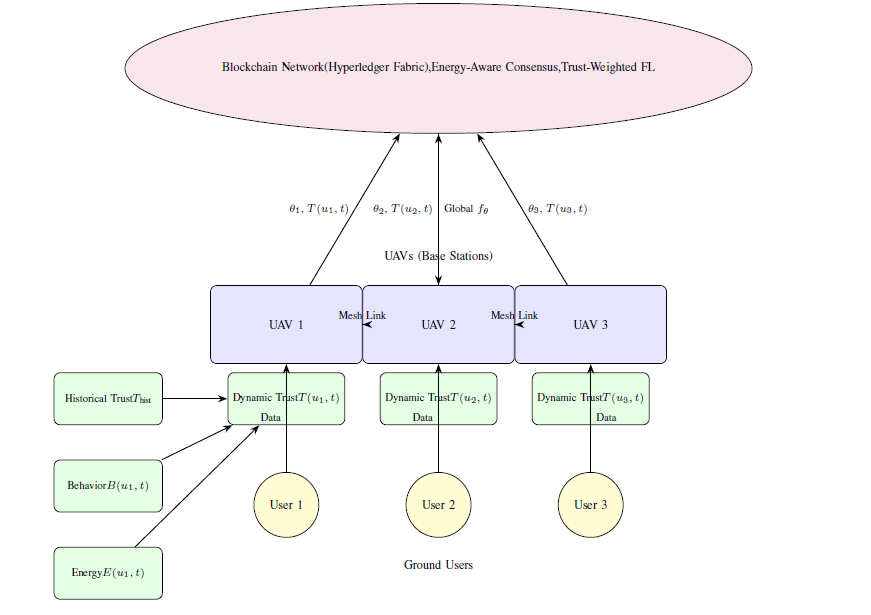}}
\caption{Proposed method illustrating UAVs in an NTN, interacting with ground users and a blockchain network with dynamic trust adjustment, energy-aware consensus, and trust-weighted FL aggregation.}
\label{fig_1}
\end{figure*}

We propose a novel Dynamic Trust Score Adjustment Mechanism with Energy-Aware Consensus (DTSAM-EAC) to protect the UAV-based NTNs ecosystem. The proposed method is depicted in Fig. \ref{fig_1}, showing the
interaction between ground users, UAVs, and the blockchain
network with the proposed dynamic trust adjustment and
energy-aware consensus mechanisms. Our approach builds upon our trust-based access control mechanism with dynamic trust score updating, energy-aware consensus protocol, and trust-weighted Federated Learning (FL) aggregation to satisfy 6G's energy efficiency, low latency, and distributed intelligence demands.

\subsection{Dynamic Trust Score Adjustment}
We introduce a dynamic trust score updating mechanism where each UAV's trust score is refreshed based on its previous trust, recent behavior, and energy contribution. Let \( T(u_i, t) \) be the trust score of UAV \( u_i \) at time \( t \), computed as:
\[
T(u_i, t) = \alpha \cdot T_{\text{hist}}(u_i, t-1) + \beta \cdot B(u_i, t) + \gamma \cdot E(u_i, t),
\]
where \( \alpha + \beta + \gamma = 1 \), and \( \alpha, \beta, \gamma \in [0, 1] \) are weighting factors. Here, \( T_{\text{hist}}(u_i, t-1) \in [0, 1] \) is the historical trust score, \( B(u_i, t) \in [0, 1] \) is the recent behavior score, and \( E(u_i, t) \in [0, 1] \) is the energy contribution score.

The behavior score \( B(u_i, t) \) is calculated as:
\[
B(u_i, t) = w_1 \cdot \text{PDR}(u_i, t) + w_2 \cdot \left(1 - \frac{\text{RT}(u_i, t)}{\text{RT}_{\text{max}}}\right),
\]
where \( \text{PDR}(u_i, t) \) is the packet delivery ratio, \( \text{RT}(u_i, t) \) is the response time, \( \text{RT}_{\text{max}} \) is the maximum acceptable response time, and \( w_1 + w_2 = 1 \).

The energy contribution score \( E(u_i, t) \) is:
\[
E(u_i, t) = \frac{\text{Energy}_{\text{remaining}}(u_i, t)}{\text{Energy}_{\text{max}}},
\]
where \( \text{Energy}_{\text{remaining}}(u_i, t) \) is the remaining battery level, and \( \text{Energy}_{\text{max}} \) is the maximum battery capacity.

\subsection{Energy-Aware Consensus Protocol}
To address energy constraints, we propose an energy-aware consensus protocol in Hyperledger Fabric. The probability \( P_{\text{validate}}(u_i, t) \) of UAV \( u_i \) being selected for block validation at time \( t \) is:
\[
P_{\text{validate}}(u_i, t) = \frac{T(u_i, t) \cdot E(u_i, t)}{\sum_{j=1}^n T(u_j, t) \cdot E(u_j, t)},
\]
where \( n \) is the number of UAVs. This prioritizes UAVs with higher trust and energy levels, reducing the energy burden on low-battery UAVs.

\subsection{Trust-Weighted FL Aggregation}
In the FL process, the global model parameters \( \theta \) are aggregated using trust scores as weights:
\[
\theta = \sum_{i=1}^n \frac{T(u_i, t) \cdot |D_i|}{\sum_{j=1}^n T(u_j, t) \cdot |D_j|} \theta_i,
\]
where \( \theta_i \) are the local model parameters, and \( |D_i| \) is the size of UAV \( u_i \)'s dataset. This ensures that more trustworthy UAVs have a greater influence on the global model.

Our approach is innovative because it introduces a dynamic update of trust score with learning in UAV behavior, unlike static trust models in \cite{b8}. Moreover, the energy-efficient consensus algorithm reduces resource utilization, which is the void of NTN security through the blockchain in \cite{b9}. Trust-weighted FL aggregation improves the credibility of the global trust model, an emerging trend of NTNs.

\section{Simulation Results}

We evaluated the proposed DTSAM-EAC Mechanism with Energy-Aware Consensus in a simulated NTN environment with 50 UAVs and 100 ground users. The simulation was carried out using NS-3 for network dynamics and Hyperledger Caliper for blockchain performance analysis. UAVs were deployed in a 5 km \(\times\) 5 km area with varying densities (sparse to dense) and topologies (mesh and star). The FL process was implemented using TensorFlow Federated, with each UAV training a local trust model (logistic regression) on interaction data (e.g., packet delivery ratio, response time). The blockchain was a permissioned Hyperledger Fabric network with the proposed energy-aware consensus protocol.

\subsection{Baselines and Metrics}
The proposed framework was compared with two baselines:

\begin{itemize}
    \item \textbf{Centralized Trust Evaluation (CTE)}: The data is sent to the central server to compute trust but not through blockchain or FL.
    \item \textbf{Standard Blockchain with Static Trust (SBST)}: A Hyperledger Fabric-based blockchain solution using static trust and scheduled consensus.
\end{itemize}

The following were compared:
\begin{itemize}
\item Prediction Accuracy of Trust Score (\%): Percentage of correctly predicted trustworthy and rogue UAVs.
\item Communication Overhead (MB per UAV): Total amount of data communicated per UAV in order to compute trust.
\item Energy Consumption (Joules per Transaction): Energy consumption incurred per blockchain transaction.
\item Convergence Time (Iterations): Number of FL iterations required by global trust model in order to converge.
\item Rogue UAV Detection Rate (\%): Percentage of correctly detected rogue UAVs
\end{itemize}

\subsection{Results and Analysis}
The simulation results are summarized in Table~\ref{tab:results}.

\begin{table}[htbp]
    \centering
    \caption{Performance Comparison of Proposed Framework Against Baselines}
    \label{tab:results}
    \begin{tabular}{l|c|c|c}
        \hline
        \textbf{Metric} & \textbf{Proposed} & \textbf{CTE} & \textbf{SBST} \\
        \hline
        Accuracy in Trust \\Score Prediction (\%) & 94 & 90 & 85 \\
        \hline
        Communication \\Overhead (MB per UAV) & 0.6 & 12 & 1.2 \\
        \hline
        Energy Consumption\\ (Joules per Transaction) & 0.3 & 0.8 & 0.5 \\
        \hline
        Convergence Time \\(Iterations) & 8 & N/A & N/A \\
        \hline
        Rogue UAV \\Detection Rate (\%) & 96 & 88 & 82 \\
        \hline
    \end{tabular}
\end{table}

 Our strategy (DTSAM-EAC) performed better than CTE (90\%) and SBST (85\%) at 94\% accuracy. FL aggregation using trust-weighted sensitivity and dynamic update of the trust score increased responsiveness to changing UAV activity. Moreover, DTSAM-EAC had an overhead of 0.6 MB per UAV, while CTE had an overhead of 12 MB (because of raw data transmission) and SBST had an overhead of 1.2 MB (because of the periodic updates of the blockchain).

\begin{figure}[htbp]
\centerline{\includegraphics[width=8cm]{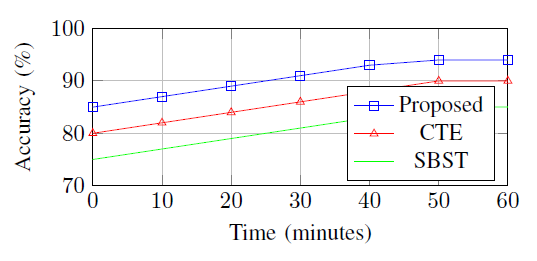}}
\caption{Accuracy in Trust Score Prediction over time.}
\label{plot_1}
\end{figure}
 
 Our proposed framework took 0.3 Joules per transaction, whereas CTE took 0.8 Joules and SBST took 0.5 Joules, as the energy-efficient consensus protocol was employed. The DTSAM-EAC framework in our paper converged in 8 iterations, i.e., efficient FL aggregation. This parameter is irrelevant to CTE and SBST.
Our system detected 96\% of the malicious UAVs, whereas CTE and SBST detected 88\% and 82\%, respectively, through dynamic trust adjustments.

\begin{figure}[htbp]
\centerline{\includegraphics[width=8cm]{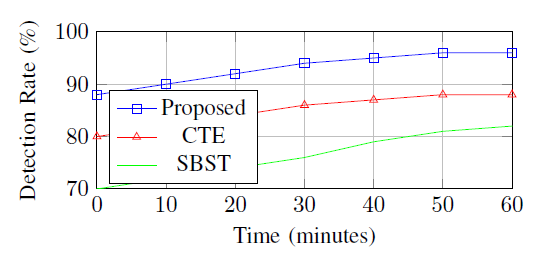}}
\caption{Rogue UAV Detection Rate over time.}
\label{plot_2}
\end{figure}

  Fig.~\ref{plot_1} and  Fig.~\ref{plot_2} reveals that the proposed framework (DTSAM-EAC) improves with accuracy as well as the rogue UAV detection rate  systematically with time, due to dynamic trust level adjustment. The proposed method outperforms CTE and SBST, which get flattened due to their static nature. Results confirm the superiority of the proposed framework with respect to privacy (lower communication overhead), energy efficiency, as well as reliability (higher accuracy as well as the detection rate), making it well-suited to 6G NTNs.

\begin{figure}[htbp]
\centerline{\includegraphics[width=8cm]{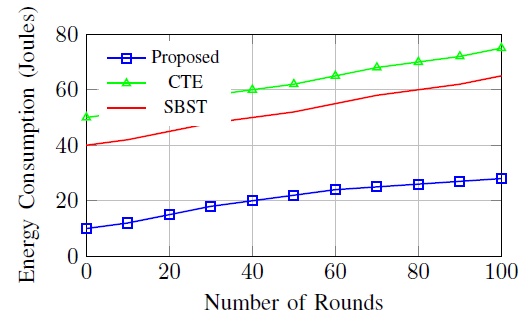}}
\caption{Energy consumption over 100 rounds for the proposed method compared to CTE and SBST.}
\label{plot_3}
\end{figure}

Fig.~\ref{plot_3} illustrates the energy expenditure (in Joules) in 100 rounds for our proposed (DTSAM-EAC), CTE, and SBST in a UAV-based NTN. DTSAM-EAC starts at 10 Joules and reaches up to 28 Joules, indicating maximum energy conservation with its energy-conscious consensus process. In contrast, CTE increases from 50 Joules to 75 Joules and SBST increases from 40 Joules to 65 Joules, due to higher energy demands from centralized communications and lack of optimization, respectively. The graph emphasizes the energy efficiency of DTSAM-EAC, which is an extremely critical aspect in resource-constrained UAVs.

\begin{figure}[htbp]
\centerline{\includegraphics[width=8cm]{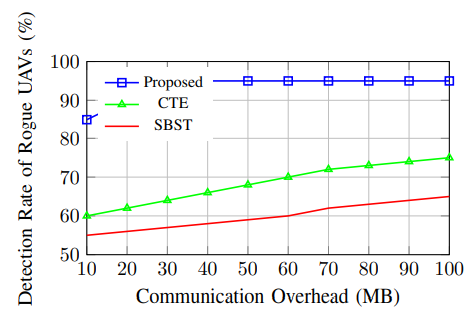}}
\caption{Detection rate of rogue UAVs Vs Communication overhead for proposed, CTE and SBST.}
\label{plot_4}
\end{figure}

Fig.~\ref{plot_4} shows the balance of DTSAM-EAC's performance in maintaining an efficient trade-off between communication efficiency and detection rate. DTSAM-EAC's jump to 95\% detection rate at just 40 MB is a proof to how well its blockchain-based trust-weighted aggregation saves data transmission and yet effectively detects malicious UAVs. Its plateau after 40 MB indicates that DTSAM-EAC is communication-efficient, without sacrificing much overhead. By contrast, SBST and CTE require much more overhead (100 MB) to yield lower detection rates (65\% and 75\%, respectively), due to centralized communication and static trust ratings that are not immune to dynamic attacks. This comparison demonstrates DTSAM-EAC's robustness in resource-constrained NTNs, where minimizing communication overhead is the secret to scalability and energy efficiency.

\section{Conclusion}

The proposed paper presented a new Dynamic Trust Score Adjustment Mechanism (DTSAM) with Energy-Aware Consensus that enhances security in UAV-based NTN. With the assistance of a Hyperledger Fabric blockchain and FL, trust scores are recurred dynamically on the basis of historic trust, present behavior, and energy contribution, and energy efficiency is optimized by the energy-aware consensus protocol. Trust-weighted FL aggregation also enhances trustworthiness in the world trust system. Simulation results indicate outstanding performance gains with 94\% prediction accuracy of the trust value and 96\% ratio of rogue UAV detection, along with enhanced privacy, energy efficiency, and trustworthiness compared to centralized and static trust-based baselines. The method is also compliant with 6G requirements of distributed intelligence and sustainability and offers a secure and scalable approach to NTNs. Deployment for real-world applications and utilization with 6G technologies like terahertz communication is foreseen for future directions for increased performance.

\end{document}